\documentclass[fleqn,12pt]{SelfArx} % Document font size and equations flushed left

\usepackage[english]{babel} % Specify a different language here - english by default
\usepackage{lipsum} % Required to insert dummy text. To be removed otherwise

\usepackage[utf8]{inputenc} % allow utf-8 input
\usepackage[T1]{fontenc}    % use 8-bit T1 fonts
\usepackage{hyperref}       % hyperlinks
\usepackage{url}            % simple URL typesetting
\usepackage{booktabs}       % professional-quality tables
\usepackage{amsfonts}       % blackboard math symbols
\usepackage{nicefrac}       % compact symbols for 1/2, etc.
\usepackage{microtype}      % microtypography
\usepackage{lipsum}	        % Can be removed after putting your text content
\usepackage[super,compress]{cite}
\usepackage{graphicx}% Include figure files
\usepackage{dcolumn}% Align table columns on decimal point
\usepackage{bm}% bold math
\usepackage{array}
\usepackage{tabularx}
\usepackage[english]{babel}
\usepackage[nottoc]{tocbibind}
\usepackage{amsmath}
\usepackage{threeparttable}
\usepackage{booktabs,caption}

\definecolor{color1}{RGB}{0,0,90} % Color of the article title and sections
\definecolor{color2}{RGB}{0,20,20} % Color of the boxes behind the abstract and headings

\usepackage{hyperref} % Required for hyperlinks
\hypersetup{hidelinks,colorlinks,breaklinks=true,urlcolor=color2,citecolor=color1,linkcolor=color1,bookmarksopen=false,pdftitle={Title},pdfauthor={Author}}

\JournalInfo{Arxive} % Journal information
\Archive{Preprint} % Additional notes (e.g. copyright, DOI, review/research article)

\PaperTitle{Period change of binary pulsars due to the accretion of dark matter particles} % Article title

\Authors{Ebrahim Hassani \textsuperscript{1}*, Hossein Ebadi \textsuperscript{2}, Reza Pazhouhesh \textsuperscript{3}} % Authors
\affiliation{\textsuperscript{1}\textit{Department of Physics, University of Birjand, Iran}} % Author affiliation
\affiliation{*\textbf{Corresponding author}: ebrahim.hassani@birjand.ac.ir}

\affiliation{\textsuperscript{2}\textit{Department of Theoretical Physics and Astrophysics, \\ University of Tabriz, PO Box 51664, Tabriz, Iran, e-mail: hosseinebadi@tabrizu.ac.ir}} % Author affiliation

\affiliation{\textsuperscript{3}\textit{Department of Physics, Faculty of Sciences, \\ University of Birjand, Birjand, Iran, e-mail: rpazhouhesh@birjand.ac.ir}} % Author affiliation

\newcommand{\keywordname}{Keywords} % Defines the keywords heading name

\Abstract{As binary systems move inside galaxies, they interact with the dark matter halo. This interaction leads to the accretion of dark matter particles inside binary components. The accretion of dark matter particles increases the mass of the binary components and then, the total mass of the system. Increased mass by this way can affect other physical parameters of the systems, like orbital periods of the systems. In this work, we estimated the period change of some known compact binary systems due to the accretion of dark matter particles inside them. Our conclustion is that, for WIMP particles with masses in the range $\simeq 10 \: GeV.c^{-2}$ and dark matter density as high as the dark matter density around the sun, period change due to accretion of dark matter partices is negligible compared to the period change due to the emission of gravitaional waves from the systems.}

\begin{document}

\flushbottom % Makes all text pages the same height

\maketitle
\tableofcontents
\thispagestyle{empty}

\section{Introduction} \label{Introduction} % The \section*{} command stops section numbering
Rotation curves of galaxies reveal the non-uniform distribution of dark matter (DM) inside galaxies \cite{Sofue2001}. From this point, one can infer that all stars that are distribited inside galaxies are immersed inside DM. Then, it is logical to suppose that the DM must affect structure and evolutionary courses of stars. For the first time Steigman used DM supposition to solve the solar neutrino problem \cite{1978AJ.....83.1050S} . Simulation of dwarf galaxies also supports the interaction between DM and stars \cite{Read_2018fxs} . In addition, DM effect on stars can be used to solve the paradox of youth problem for stars that are located near the Galactic massive black hole \cite{Hassani_2020zvz} . In addition to normal stars, DM effect on other celestiall bodies like the moon \cite{2020PhRvD.102b3024C, 2020PhLB..80435403G} , planets \cite{2012JCAP...07..046H} , neutron stars (NS) \cite{2012PhRvD..85b3519M, 2011IJMPE..20..109R, 2010PhRvL.105n1101P, 2010PhRvD..81l3521D, Bertone_2008, 2008PhRvD..77b3006K, Bell:2020jou, Joglekar:2020liw, 2020arXiv200405312L, 2019PhRvD.100j3019H, 2019JCAP...06..054B, 2020JCAP...06..007F, 2018JCAP...09..018B, 2018JHEP...11..096K, 2017PASA...34...43C, 2017arXiv170508864H, 2018ApJ...863..157C, 2017PhRvL.119m1801B, Perez-Garcia:2015rta, Perez-Garcia:2014nza, 2015ApJ...800..141Z, Tachibana:2013iva, 2013arXiv1306.0148P, Perez-Garcia:2013dwa, 2013PhRvD..87l3507B, 2013PhRvD..87e5012B, Serebrov:2012vm, 2012AIPC.1441..525P} , white dwarf stars (WD) \cite{Wong:2011ege, 2019JCAP...08..018D, 2018PhRvD..98f3002C, 2018PhRvD..98k5027G, 2018PhRvD..98j3023N, 2017arXiv170310104N, 2016MNRAS.459..695A, 2015PhRvD..91j3514H, 2013PhRvD..87l3506L}, black holes \cite{2020arXiv200405312L, 2012PhRvD..85b3519M, 2009JCAP...08..024U, Belotsky2014} and binary star systens \cite{Hassani2020} are also investigated in the literature. \\
For the first time, Press and Spergel estimated the rate at which DM particles (WIMP particles, exactly speaking) accrete into the sun \cite{1985ApJ...296..679P} . In the next step, Gould obtained a general relation for capture rate (CR) of DM particles by other round bodies like planets and stars \cite{Gould_1987ir} . Kouvaris used Press and Spergel relation to derive CR relation for NSs \cite{2008PhRvD..77b3006K} (see section \ref{Sec_CR_by_NS} for more details). Hurst et al used Gould relation to obtain CR relation for WDs \cite{2015PhRvD..91j3514H} (see section \ref{Sec_CR_bY_WDs} for more details).\\
As a compact binary system moves inside the DM halo, it feels dynamical friction by the DM halo. Induced dynamical friction by the DM halo can alter the orbital parameters (e.g. period and semi-major axis) of these systems. In a series of studies, reserchers tried to estimate the period change of pulsar binary systems due to this dynamical friction and compare the results with the period change due to the gravitational wave emission \cite{PhysRevD.96.063001, Penarrubia2016, Blas2020, GOMEZ2019100343, Blas2017, 2015PhRvD..92l3530P, Armaleo2020,  CAPUTO20181, Yoo2004}.\\
In this work we estimated the period change of compact binary systems due to the accretion of DM partilces (instead of due to the dynamical friction) inside them. Accretion of DM particles inside compact binary systems increases the mass of the binary components. Then, increased mass leads to the change in the orbital periods of these system. To the best of my knowledge, there is not any previous study that estimates the period change of binary systems due to the accretion of DM particles.\\
The rest of this paper is structured as follows. In sections \ref{Sec_CR_bY_WDs} and \ref{Sec_CR_by_NS} we presented CR ralations for WDs and NSs, separately. In section \ref{Sec_CR_By_Compact_Binary_Systems} we presented an approximate relation for the CR by compact binary systems. Finally, section \ref{Sec_Results_and_discussion} devoted to the results and discussions.

\section{Capture rate by white dwarfs} \label{Sec_CR_bY_WDs}
Capture Rate of DM particles by white dwarfs can be calculated using the equation \cite{2015PhRvD..91j3514H} :
\begin{equation} \label{Eq_CR_by_WDs}
CR_{WD} =  \underset{i}{\sum} CR_{WD, i} =\sqrt{\frac{3}{2}} \: \frac{\rho_{x}}{m_{x}} \: \sigma_{\chi, i} \: \frac{v_{esc,R_{WD}}}{\bar{v}} \: N_{i} \: \left \langle \widehat{\phi } \right \rangle \: \frac{erf(\eta)}{\eta }
\end{equation}
like equation \ref{Eq_CR_NS} for neutron stars, $ \rho_{\chi} $, $ m_{\chi} $ and $ \bar{v} $ are the density of DM halo around white dwarf (or around compact binary system), mass of WIPM particles and velocity dispersion of WIMPs (which we supposed to be $ \bar{v} = 220 \: km.sec^{-1}$ for all stars in this study), respectevely. $ R_{WD} $ is the radius of the white dwarf and $ v_{esc,R_{WD}} $ is the escape velocity at the surface of the white dwarf. $ \eta $ is the ratio of star's velocity through its halo to the local velocity dispersion of the DM halo. \\
$ \sigma_{\chi, i} $ is the scattering cross section of WIMP particles from nuclear species $i$ . As WDs consumed almost all of their hydrogen then, we just consider the spin-independent part of the scattering cross section and neglect the spin-dependent part. So, the spin-independent scattering cross section for all elements except hydrogen is \cite{1996PhR...267..195J, 2015PhRvD..91j3514H, Lopes2011} :
\begin{equation} \label{Eq_Scatterin_cross_wd}
\sigma_{\chi ,i} = \sigma_{\chi ,SI} A_{i}^{2}\left ( \frac{m_{\chi} m_{n, i}}{m_{\chi}+m_{n, i}} \right )^{2} \left ( \frac{m_{\chi}+m_{proton}}{m_{\chi}m_{proton}} \right )^{2}
\end{equation}
where $ \sigma_{\chi ,SI} $ is the spin-independent scattering cross-section which is determined through experimental direct DM detection methods. In this study we considered it to be $ \sigma_{\chi, SI} = 10^{-44} \: cm^{2} $ (as mentioned again in the section \ref{Sec_CR_by_NS} too). $ A_{i} $ is atomic number of species $i$, $ m_{n, i} $ in nuclear mass of species $i$ and $m_{proton}$ is the mass of a proton. To simplify calculations, we supposed that WDs composed of just carbon atoms with $m_{n,c} = 12 \: u$ where $u$ is the atomic mass unit (amu) and atomic number $A_{c} = 6$ . The similar supposition had done in the similar work \cite{2015PhRvD..91j3514H} . \\
In equation \ref{Eq_CR_by_WDs}, $ \widehat{\phi } $ is dimensionless potential for stellar nucleons and defined as \cite{2015PhRvD..91j3514H} :
\begin{equation}
\widehat{\phi } = \dfrac{v_{esc}^{2}(r)}{v_{esc}^{2}(R)}
\end{equation}
The quantity $\left \langle \widehat{\phi } \right \rangle$ is the average of $\widehat{\phi}$ over all necleons in the star. To calculate $\left \langle \widehat{\phi } \right \rangle$, we supposed that, like the case for NSs, average density of all WDs is constant. So, for a typical WD (e.g. Sirius B WD star with $M_{Sirius B} = 1.018 \: M_{\odot}$ mass \cite{Bond2017} and radius \cite{Holberg1998} $M_{Sirius B} = 0.0084 \: R_{\odot}$) we have:
\begin{equation} 
\bar{\rho}_{WD} = \dfrac{M_{WD}}{(4/3) \pi R^{3}_{WD}} = \dfrac{1.018 \: M_{\odot}}{(4/3) \pi (0.0084 \: R_{\odot})^{3}} = 2.38 \times 10^{9} \: \: \: (\dfrac{Kg}{m^{3}})
\end{equation}
So, the approximate mass-radius relation for white dwarfs becomes :
\begin{equation}
R_{WD} = \left(  \dfrac{M_{WD}}{(4/3) \pi \bar{\rho}_{WD}} \right) ^{\dfrac{1}{3}} = 4.65 \times 10^{-4} \times M_{WD}^{1/3} \: \: \: (m) .
\end{equation}
(We need $R_{WD}$ to calculate $v_{esc}(R)$ for WDs with different masses too). Now, $\left \langle \widehat{\phi } \right \rangle$ becomes:

\begin{equation}
\left \langle \widehat{\phi } \right \rangle = \dfrac{1}{R} \int_{0}^{R} \frac{v_{esc}^{2}(r)}{v_{esc}^{2}(R)}dr = \frac{1}{3}
\end{equation}
In equation \ref{Eq_CR_by_WDs}, $N_{i}$ is the total number of nucleons of species $i$ in the WD star. In the case of carbon WDs it becomes:
\begin{equation}
N_{c} = \dfrac{M_{WD}}{m_{c}} = \dfrac{M_{WD}}{12 \: u}
\end{equation}

\section{Capture rate by neutron stars} \label{Sec_CR_by_NS}
Capture Rate by Neutron stars, considering the general relativity corrections, can be calculated using the below equation \cite{2008PhRvD..77b3006K} :
\begin{equation} \label{Eq_CR_NS}
CR_{NS} = 4 \pi^{2} \: \frac{\rho_{x}}{m_x} \: (\frac{3}{2 \pi \bar{v}^{2}})^{3/2} \: \: \: \frac{(2GM_{NS}R_{NS})}{1-\frac{2GM_{NS}}{c^{2} \:R_{NS}}} \times min(\frac{1}{3} \: \bar{v}^{2} , E_{0}) \times f
\end{equation}

where $\rho_{x}$ is the density of the DM halo around the neutron star, $m_{x}$ is the mass of the WIMP particles, $ \bar{v} $ is the the velocity dispersion of the DM particles (in this study we took it to be $ \bar{v} = 220 \: km.sec^{-1}$ for all stars in the study), G is the gravitational constant, c is the speed of light, $M_{NS}$ is the mass of the neutron star and $R_{NS}$ is the radius of the neutron star. \\
To estimate the radius of a neutron stars $R_{NS}$ (to use in equation \ref{Eq_CR_NS}), we supposed that the average density of the all neutron stars is constant. So, the average density of a typical neutron star with $ M_{NS} = 1.44 \: M_{\odot} $ and radius $ R_{NS} = 10.6 \: km $ (see table 1 of the paper \cite{2014JCAP...05..013G} for these quantities) is :

\begin{equation} 
\bar{\rho}_{NS} = \dfrac{M_{NS}}{(4/3) \pi R^{3}_{NS}} = \dfrac{1.44 \: M_{\odot}}{(4/3) \pi (10.6 \times 10^{3})^{3}} = 5.74 \times 10^{17} \: \: \: (\dfrac{Kg}{m^{3}})
\end{equation}
So, we obtain an approximate mass-radius relation for neutron stars:
\begin{equation} 
R_{NS} = \left(  \dfrac{M_{NS}}{(4/3) \pi \bar{\rho}_{NS}} \right) ^{\dfrac{1}{3}} = 7.46 \times 10^{-7} \times M_{NS}^{1/3} \: \: \: (m) .
\end{equation}
In equation \ref{Eq_CR_NS}, $ E_{0} $ is a constant that parametrizes the maximum kinetic energy per mass of the WIMPs at an asymptotically large distance from the star in order for the WIMPS to be captured by the star. In the case of neutron stars and WIMPs with masses of the order GeV, $ E_{0} \gg (1/3) \bar{v}^{2}$. Then, in equation \ref{Eq_CR_NS}, $ (1/3) \bar{v}^{2} $ should be taken always as minimum (for more detailded dicussion about this selection see the section 2 of the paper \cite{2008PhRvD..77b3006K} .

Not all the WIPMs that enter a neutron star will be captured by it. In equation \ref{Eq_CR_NS}, $f$ represents the fraction of the WIPMs that, after one or several scattering from neclear matter, lose enough energy to be captured by a typical neutron star. For WIMP-necleon scattering cross sections $ \sigma_{\chi} > 10^{-45} \: cm^{2} $, $f = 1 $ (as in the case of this study) ; otherwise $f$ is given by \cite{2008PhRvD..77b3006K} :
\begin{equation} 
f = 0.45  \dfrac{\sigma_{\chi}}{\sigma_{crit}}
\end{equation}
which $ \sigma_{crit} = m_{n} R^{2} / M \simeq 6 \times 10^{-46} cm^{2} $. In this study we supposed $ \sigma_{\chi} = 10^{-44} cm^{2} $ (then, $f = 1$) which is the maximum magnitude that are determined through the experimental DM detection experiment \cite{2011PhRvL.106b1303B}.

\section{Capture rate by compact binary systems} \label{Sec_CR_By_Compact_Binary_Systems}
According to the Kepler's third law, square of the period in binary systems is proportional to the cube of the semi-major axis of the orbit \cite{hilditch_2001} :
\begin{equation} \label{Eq_Keplers_third_law}
P^{2} = (\dfrac{4 \pi^{2}}{GM}) \: a^{3} 
\end{equation}
which $M=M_{1}+M_{2}$ . If we suppose that all binary systems that are located inside the galaxies immeresed inside the DM halos then, it is logical to suppose that the both components of the systems accumulate DM particles while they are orbiting around each other. By this way, by passing the time, the total mass of the systems will increase. In the next, we will estimate how much the incresed mass will affect the other parameters of the systems. Taking differential of the both sides of the equation \ref{Eq_Keplers_third_law} yealds:
\begin{equation}
\begin{split}
2 P dP = \dfrac{4 \pi^{2}}{G} d(\dfrac{a^3}{M})  \\ 
= \dfrac{4 \pi^{2}}{G} \dfrac{3a^{2}da}{(M_{1}+M_{2})} - \dfrac{4 \pi^{2}a^{3}}{G} \dfrac{dM_{1}}{(M_{1}+M_{2})^{2}} - \dfrac{4 \pi^{2}a^{3}}{G} \dfrac{dM_{2}}{(M_{1}+M_{2})^{2}} \\
= \underbrace{\frac{4 \pi^{2} a^{3}}{G(M_{1}+M_{2})}}_{=P^{2}} \left [ \frac{3}{a}da - \frac{dM_{1}+dM_{2}}{M_{1}+M_{2}} \right ]
\end{split}
\end{equation}
which simplifies to :
\begin{equation} \label{Eq_Kepler_law}
\begin{split}
\dfrac{dP}{P} = \dfrac{3}{2} \dfrac{da}{a} - \dfrac{dM}{2M} = \dfrac{3}{2} \dfrac{da}{a} - \dfrac{dM_{1}+dM_{2}}{2(M_{1}+M_{2})}
\end{split}
\end{equation}
multiplying both sides of the equation \ref{Eq_Kepler_law} to $1/dt$ yealds :
\begin{equation} \label{Eq_CR_by_binary_systems}
\frac{\dot{P}}{P} = \frac{3}{2} \frac{\dot{a}}{a} - \frac{1}{2} \frac{\dot{M}}{M} = \frac{3}{2} \frac{\dot{a}}{a} - \frac{1}{2} \frac{\dot{M_{1}}+\dot{M_{2}}}{(M_{1} + M_{2})}
\end{equation}
where $\dot{P} = dp/dt$, $\dot{a} = da/dt$, $\dot{M} = dM/dt$, $\dot{M_{1}} = dM_{1}/dt$ and $\dot{M_{2}} = dM_{2}/dt$. $\dot{M_{1}}$ and $\dot{M_{2}}$ signifies the accretion of DM particles into compact stars. To calculate $\dot{M_{1}}$ and $\dot{M_{2}}$ it is enough to multiply equation \ref{Eq_CR_NS} (in the case of neutron stars) or equation \ref{Eq_CR_by_WDs} (in the case of white dwarf stars) to the mass of the DM particles $m_{\chi}$ (e.g. $\dot{M_{1}} = CR_{NS} \times m_{\chi}$ if the $M_{1}$ star is a neutron star). In table \ref{table_CR_in_known_binary_systems} we estimated  $\dot{M} / M $ (which $\dot{M} = \dot{M}_{1} + \dot{M}_{2}$) for some known compact binary systems. \\

\section{Results and Conclusions} \label{Sec_Results_and_discussion}
Using equation \ref{Eq_CR_by_binary_systems}, it is possible to estimate the accretion of DM particles in compact binary systems. 
\begin{table}
\centering
\caption{Compact binary system parameters that we used in this study to test our results. All parameters in the table are obtained through observations, except the $\dot{M}/M$ column which is obtained theoretically using the assumptions that are explained in the text.}
\begin{threeparttable}
\begin{tabular}{|| c | c || c | c || c | c | c | c | c | c ||}
\hline \hline
Name & \shortstack{ $M_{1}$ \\ $(M_{\odot})$ } & \shortstack{ $M_{2}$ \\ $(M_{\odot})$}  & \shortstack{ $ P_{d} $ \\ $(days)$ } & $\dot{P}$ \tnote{*} & \shortstack{ $\dot{P}/P$ \\ $(sec^{-1})$} & \shortstack{ $\dot{M}/M$ \\ $(sec^{-1})$} &  $Ref.$ \tnote{**} \\ \hline 

J1903+0327  & 1.0   & 1.7 & 95   & -6.4e-11 &  -7.8e-18  & 1.9e-32 & \cite{Freire2011} \\
J0737-3039  & 1.2   & 1.3 & 0.10 & -4.0e-15 &  -4.6e-19  & 3.5e-45 & \cite{Wex2014} \\
B1913+16    & 1.4   & 1.4 & 0.32 & 5.0e-15  &  1.8e-19   & 3.6e-45 & \cite{Wex2014} \\
J1012+5307  & 0.16  & 1.6 & 0.60 & -1.8e-14 &  -3.5e-19  & 2.5e-32 & \cite{Arzoumanian2018, Desvignes2016} \\
J1614-2230  & 0.49  & 1.9 & 8.7  & 3.4e-13  &  4.5e-19   & 2.6e-32 & \cite{Arzoumanian2018} \\
J1909-3744  & 0.21  & 1.5 & 1.5  & -4.0e-15 &  -3.1e-20  & 2.3e-32 & \cite{Arzoumanian2018, Desvignes2016} \\
J0751+1807  & 0.16  & 1.6 & 0.26 & -4.6e-14 &  -2.1e-18  & 2.5e-32 & \cite{Desvignes2016} \\
J1910+1256  & 0.19  & 1.6 & 58   & -2.0e-11 &  -4.0e-18  & 2.5e-32 & \cite{Gonzalez2011} \\
J2016+1948  & 0.29  & 1.0 & 635  & -1.0e-9  &  -1.8e-17  & 2.7e-45 & \cite{Gonzalez2011} \\
J0348+0432  & 0.17  & 2.0 & 0.10 & -1.1e-14 &  -1.3e-18  & 3.2e-32 & \cite{2013Sci...340..448A} \\
J1713+0747  & 0.29  & 1.3 & 68   & 3.0e-14  &  5.1e-21   & 2.9e-45 & \cite{Zhu2019} \\
J0613-0200  & 0.12  & 1.2 & 1.2  & 2.7e-14  &  2.6e-19   & 2.5e-45 & \cite{Arzoumanian2018, Desvignes2016} \\
J1738+0333  & 0.19  & 1.5 & 0.35 & 2.0e-15  &  6.6e-20   & 2.4e-32 & \cite{2012MNRAS.423.3328F} \\ \hline

\end{tabular}
\begin{tablenotes}
\item[*] $\dot{P}$ is a dimensionless quantity.
\item[**] References for all parameters except $\dot{M}/M$, which is calculated in this study.
\end{tablenotes}
\end{threeparttable}
\label{table_CR_in_known_binary_systems}
\end{table}
In our calculations, we supposed stars with masses lower than the chandrasekhar mass ($1.4 \: M_{\odot}$) to be white dwarfs \cite{2000itss.book.....P} . Stars with masses higher than $1.4 \: M_{\odot}$ considered to be neutron stars. The mass of the WIMP particles and density of DM halo around binary systems considered to be $m_{\chi}=10 \: \: GeV.c^{-2}$ and $\rho_{\chi} = 8.16 \times 10^{-3} M_{\odot}/pc^{3}$ (which is the estimated DM density around solar neibourhoud \cite{Bertone2005a} ), respectively. \\
In equation \ref{Eq_CR_by_binary_systems}, it is possible to measure $\dot{P}/P$ term through observations.  But the precision of the current observational technology is not enough to measure the $ \dot{a}/a $ term in the right hand side of the equation. \\
Paying attention to the amounts of $\dot{M}/M$ and $\dot{P}/P$ in table \ref{table_CR_in_known_binary_systems}, we see that the amounts of $\dot{M}/M$ are about 12-14 orders of magnitude smaller than the amounts of $\dot{P}/P$ (for binaries that at least one of the components is a neutron star). For binaries that both of the components are white dwarfs (e.g. J0737-3039, B1913+16 or J2016+1948 systems) the results are more dramatic and the amounts of $\dot{M}/M$ are about 24-26 orders of magnitude smaller than the amounts of $\dot{P}/P$. Then, we can say that the accretion of DM particles is not the main reason of period change in compact binaries. In order to interpret the period changes $\dot{P}/P$, other physical sources that affects systems parameters, e.g. gravitaional emission, mass transfer, dynamical friction etc., must be token into account. \\
As an overall result, our calculations show that the accretion of DM particles in compact binary systems do not change the systems parameters significantly. In addition, accretion of DM particles can not be considered the main source of period changes in binary pulsars.

\phantomsection
\section*{Acknowledgments} % The \section*{} command stops section numbering
\addcontentsline{toc}{section}{Acknowledgments} % Adds this section to the table of contents
The author would like to express his special thanks to Prof. Joakim Edsjö from the University of Stockholm, Sweden, and Dr. Amin Rezaei Akbarieh from the University of Tabriz, Iran and Prof. Gianfranco Bertone from the University of Amsterdam, Netherlands and Dr. Marco Taoso from the Istituto Nazionale di Fisica Nucleare (INFN), Italy because of their worthwhile discussions during the research.
\phantomsection
\bibliographystyle{ieeetr}
\bibliography{My_References}
\end{document}